\documentclass{elsart}

\usepackage{amssymb}
\usepackage{graphicx}

\begin{document}

\begin{frontmatter}



\title{Face Recognition in the Machine Reveals Properties of Human Face Recognition}


\author{Matthias S. Keil$ ^a$, \`Agata Lapedriza$ ^a$, David Masip$ ^b$ and Jordi Vitri\`a$^a$}

\address{
$^a$Centre de Visi\'o per Computador (CVC), Dept. Inform\`atica. \\Universitat Aut\`onoma de Barcelona  \\
Bellaterra, Spain, 08193. \\
 \{mats,agata, jordi\}@cvc.uab.es \\$^b$Department of Applied Mathematics and Analysis (MAiA)\\ University of Barcelona (UB)\\ Edifici
Hist\`oric Gran Via de les Corts Catalanes 585, Barcelona 08007, Spain.\\ davidm@maia.ub.es\\}

\begin{abstract}
Psychophysical studies suggest that face recognition takes place in a narrow band of low spatial
frequencies (``critical band'').  Here, we examined the recognition performance of an artificial
face recognition system as a function of the size of the input images.  Recognition performance was
quantified with three discriminability measures: Fisher Linear Discriminant Analysis, non
Parametric Discriminant Analysis, and mutual information. All of the three measures revealed a
maximum at the same image sizes.  Since spatial frequency content is a function of image size, our
data consistently predict the range of psychophysical found frequencies.  Our results therefore
support the notion that the critical band of spatial frequencies for face recognition in humans and
machines follows from inherent properties of face images.

\end{abstract}

\end{frontmatter}

\section{Introduction}
A considerable number of psychophysical studies coincide in that mechanisms for face recognition in humans
do not use all available visual information about faces equally.  The visual information we refer to concerns
the spatial frequency composition of face images.  Specifically, a narrow band settled at the lower end of the
frequency spectrum seems to be optimally suited for the recognition of previously learnt faces.  The frequency
band is centered at about $10$ to $15$ cycles per face, and its bandwidth is about $2$ octave
\cite{Ginsburg1984,TiegerGanz1979,FiorentiniMaffeiSandini1983,HayesMorroneBurr1986,PeliEtAl1994,%
CostenParkerCraw1994,Nasanen1999,OjanpaaNasanen2003}.  Thus, face recognition (and also object
recognition in general) follows a bandpass characteristics in humans.  Furthermore, this result
does not depend, to a first approximation, on viewing instance
\cite{HayesMorroneBurr1986,OjanpaaNasanen2003}.  The unit for
spatial frequencies ``cycles per face'' (or ``cycles per object'') expresses this scale invariance.\\
Nonetheless, to the best of our knowledge, no conclusive explanation for the bandpass
characteristic of face recognition has emerged so far.  A recent study conducted by one of the
authors linked this characteristic to inherent properties of face images.  By examination of the
responses of a model of simple and complex cells to face images, Keil could show that higher
response amplitudes are obtained at spatial frequencies consistent with the corresponding
psychophysical data \cite{KeilFace2006}.  Therefore, the visual system should encode visual
information for processing faces preferably at those spatial frequencies, where the highest
signal-to-noise ratio is obtained. Only then a fine discrimination between signals will be
possible.  Or, otherwise expressed, only then
we will be able to learn and perceive fine differences between otherwise similar faces.\\
Given this link between the statistics of facial images and psychophysical data, we reasoned that
an artificial face recognition system should reveal similar properties as the human visual system
does: we expected to see an optimal recognition performance of the artificial system at the same
spatial frequencies as observed with humans.  To this end, we explored several measures of
recognition performance.  Furthermore, as suggested by the results from ref. \cite{KeilFace2006},
internal face features are the principal cause for the bandpass characteristic of face recognition.
This holds especially true for the eyes, but also for mouth and nose, albeit to a less extent.
Consequently, we suppressed external features (hairline, shoulder regions) in the present study.
The results of the present study suggest that the machine indeed does it like humans -- recognition
performance peaks within a narrow band of low spatial frequencies.\\

This paper is organized as follows: the next section describes the image processing and the
separability criteria that have been considered in the experiments, section 3 shows the obtained
results, and section 4 summarizes and concludes this work.

%
%
%
%
%
%
%
%
%
%
%
%
\section{Methods}
%
%
%
\subsection{Processing of face images}\label{preparation}
\begin{figure}[th]
\centering
\includegraphics[width=2.75in]{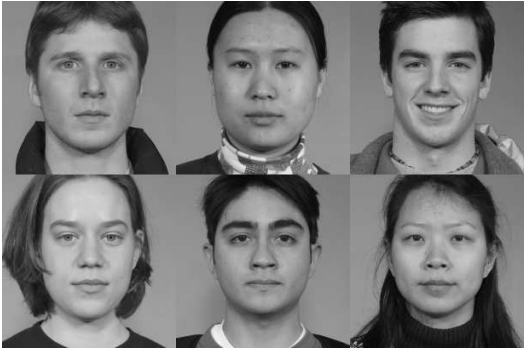}
\caption{Example images from the FRGC database.  The images were acquired
under controlled conditions.} \label{FRGC_database_samples}
\end{figure}

\begin{figure}[th]
\centering
\includegraphics[width=2.75in]{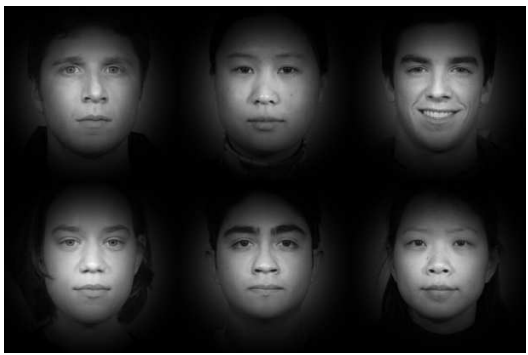}
\caption{The images shown here are the result of applying a Blackman-Harris window to the images of
Figure \ref{FRGC_database_samples}.  The application of the Blackman-Harris window leads to a good
suppression of external face features (e.g. hair and shoulders).}
\label{FRGC_database_samples_with_BH}
\end{figure}

\begin{figure}[th]
\centering
\includegraphics[width=2.75in]{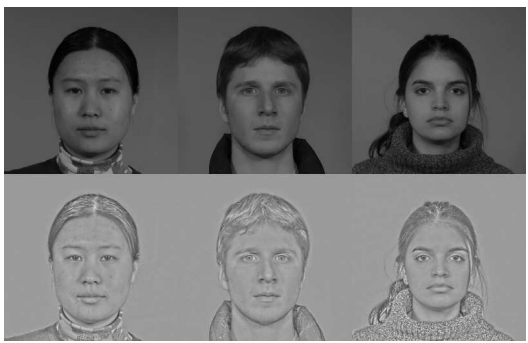}
\caption{Examples of images after applying the ``Weber-filtering''.  Compared
to the original images (top row), Weber-filtering leads to enhancement of high spatial
frequencies, and discounts illumination effects at the same time (bottom row).}
\label{WeberSamples}
\end{figure}

%
For our experiments, we used images from the FRGC Database
(\textit{http://www.bee-biometrics.org/}). In these images, faces
appear against uniform, grey background, and with homogeneous
illumination conditions for all subjects. The database consisted
of $3772$ high quality images from $275$ different persons, where
four to $32$ images exist for each subject. To perform the
experiments we first aligned and then resized the images such that
each resulting image had an eye-to-eye distance of $50$ pixels.
Figure \ref{FRGC_database_samples} shows some examples of the such
normalized images.\\
Due to their relatively high variability, we decided to suppress external face features by windowing
each face image with a 4-term Blackman-Harris window (``B.H.-window'').  This window was compared to
$15$ alternative windows, and scored the highest similarity with corresponding images whose external
features were removed manually \cite{KeilFace2006}. For each image, the window
was centered at the position of the nose $(x_\mathit{nose},y_\mathit{nose})$.  The nose position was
estimated from the coordinates of the left and the right eye $(x_\mathit{le},y_\mathit{le})$,
$(x_\mathit{re},y_\mathit{re})$, respectively, and the mouth $(x_\mathit{mouth},y_\mathit{mouth})$:
\def\rnd{\mathrm{rnd}}
\begin{eqnarray}\label{NosePosition}
      x_\mathit{nose}   &=& \rnd\left((x_\mathit{le}+x_\mathit{re})/4  + x_\mathit{mouth}/2\right)\\\nonumber
      y_\mathit{nose}   &=& \rnd [0.95*\rnd (y_\mathit{le}  + (y_\mathit{mouth}\\\nonumber
                 & & \ \ \ \ \ \ \ \ - (y_\mathit{le}+y_\mathit{re})/2)/2 ) ]
\end{eqnarray}
The operator $\rnd(\cdot)$ denotes rounding of its argument to the
nearest integer value. Figure \ref{FRGC_database_samples_with_BH}
illustrates result of applying the B.H.-window to the images of
Figure \ref{FRGC_database_samples}.\\
We adopted the following procedure to assess the frequency-dependence of face recognition with our
artificial system. Each image was down-sized to continuously smaller sizes. After down-sizing, we
enhanced the highest spatial frequencies with a modified algorithm for suppressing illumination
effects \cite{GrossBrajovic03}.  Specifically, we modified the algorithm such that its output
mimicked the responses of retinal ganglion cells \cite{Kuffler53} in a way that contour enhancement
at high spatial frequencies occurred (Figure \ref{WeberSamples}).  Consequently, our
``Weber-filtered'' images are dominated by the Nyquist frequencies associated with each image size.
In this way, computational time could be saved over naive bandpass filtering.
%
%
\subsection{Separability Measures}\label{measures}

In order to measure the optimal dimensionality we need to define a formal class separability
criterion. Class separability can be measured in terms of classification accuracy or class
distribution. In the first case, the measure is highly dependent on the classifier used. In this
paper we propose the use of a classifier-independent set of statistical
measures to validate the psychophysical results for human face recognition.\\
Two types of class separability measures are described in the literature \cite{Fukunaga90}, where
one is based on scatter matrices, and the other one is based on imposing an upper limit on the
Bayes error (Bhattacharyya distance). In this paper we will focus on the former measure (scatter
matrices), because the latter approach necessitates the estimation of probability distributions,
which is a notoriously
difficult endeavor.\\
A further statistical criterion to measure the separability between classes is based on mutual
information, which is defined as:
\begin{equation}
I(X,Y)=\int\int p(X,Y)log \left ( \frac{p(X,Y)}{p(X)p(Y)} \right ) dXdY
\end{equation}
where $X$ and $Y$ are two random variables, and $p(X)$ and $p(Y)$ their respective probability
density functions. In this paper we compute mutual information between data points $X$ and classes
$C$. A large value of mutual information in this case means that we have much information about the
class $C$ given the observation $X$.  On the other hand, if the mutual information is zero, then
both variables are independent.  Notice that the computation of mutual information also
necessitates the estimation of corresponding probability distributions.  However, Torkkola
\cite{Torkkola03} recently proposed a method which makes the computation of mutual information
feasible by using a quadratic divergence measure that allows an efficient non-parametric
implementation, without prior assumptions about class densities. More concretely, his approximation
is inspired by the quadratic Renyi entropy, and the method can be used with training data sets of
the order of tens of thousands of samples.

%
\subsection{Discriminant Analysis}\label{discriminant}
%
Classic discriminant analysis techniques have often been applied to linear feature extraction in
order to find the projection matrix that preserves the class separability of data points.
Typically, two kind of statistics have been used for this purpose: \emph{(i)} the within class
scatter matrix that shows the scatter of samples around the same class, and \emph{(ii)} the between
class scatter matrix.\\
In order to formulate a criterion for class separability, each matrix has to be reduced to a single
and unique number.  This number should be large when the between class scatter is large -- or when
the within class variation is small. Several ways for computing the number have been defined in the
literature:
\def\trace{\mathit{trace}}
\begin{eqnarray}
J_1 &=& \trace(S_2^{-1}S_1) \label{crit_1}\\
J_2 &=&ln|S_2^{-1}S_1| \label{crit_2}\\
J_3 &=&\frac{\trace(S_1)}{\trace(S_2)} \label{crit_3}
\end{eqnarray}
In the classic feature extraction literature the $J_1$ criterium is used, given that it can be
maximized using a closed formulation. The general technique to get the job done is known as Fisher
Linear Discriminant Analysis (FLD) \cite{Fisher36}, and uses as ${\bf S}_1$:
\begin{equation}
    \textbf{S}_B=\frac{1}{K}\sum_{k=1}^K ({\bf m}_k-{\bf m}_0)({\bf m}_k-{\bf m}_0)^T
    \label{eq_FLD_between}
\end{equation}
where ${\bf m}_k$ is the class-conditional sample mean and $\textbf{m}_0$ is the unconditional
(global) sample mean. Furthermore, for ${\bf S}_2$:
\begin{equation}
    \textbf{S}_W=\frac{1}{K}\sum_{k=1}^K {\bf S}_k
    \label{eq_FLD_within}
\end{equation}
where ${\bf S}_k$ is the class-conditional covariance matrix for $C_k$ estimated from the data.\\
The main problem with the classical FLD approach is that the optimization of the criterion
(\ref{crit_1}) using ${\bf S}_B$ and ${\bf S}_W$ is blind for anything beyond second order
statistics.  As a consequence, it may be inaccurate for measuring separability of more complex
structures.  To remedy, Fukunaga and Mantock \cite{Fukunaga83} propose to use a non-parametric
estimated between-class scatter matrix $\textbf{S}_B$, which has generally a full rank.  This
estimation was used in the non Parametric Discriminant Analysis algorithm (NDA), which has been
shown to considerable improve the accuracy of the classic FLD.  In a nutshell, the non parametric
between class scatter matrix is estimated as follows.\\
Let ${\bf x}$ be a data point in ${\bf X}$ with class label $C_j$, and by
$x^{\rm{\overline{class}}}$ the subset of the $k$ nearest neighbors of ${\bf x}$ among the data
points in ${\bf X}$ with class labels different from $C_j$. We calculate a local between-class
matrix for ${\bf x}$ as:
\begin{equation}
\Delta_{B}^{\bf x}=\frac{1}{k-1}\sum_{{\bf z}\in x^{\rm{\overline{class}}}} ({\bf z}-{\bf x})({\bf
z}-{\bf x})^T
\end{equation}
The estimate of the between-class scatter matrix ${\bf S}_B$ is found as the average of the local
matrices
\begin{equation}
{\bf S}_B=\frac{1}{N}\sum_{{\bf z}\in X} \Delta_B^{\bf z}
\end{equation}
The resulting ${\bf S}_B$ is used in the criterion (\ref{crit_1}) as the new $S_1$.
%
%
\section{Results}
\begin{figure}[th]
\centering
\includegraphics[width=4.6in]{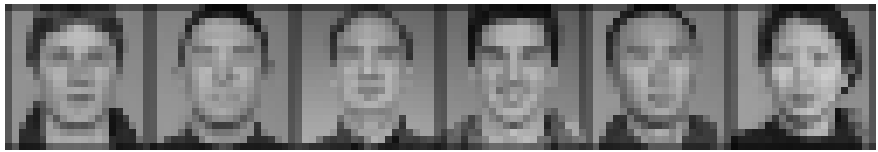}
\caption{Examples at image sizes which are associated with maximum discrimination. ($22\times22$ ).
The image sizes translate to approximately $5-8$ cycles per face width, and $6-12$ cycles per face
height, and are thus within the ballpark of the corresponding psychophysical data. Notice that here
the re-sized \emph{original} images are shown (i.e., without application of the Blackman-Harris
window) to achieve a better visibility.} \label{FRGC_database_small}
\end{figure}

\begin{figure}
\centering
\begin{tabular}{cc}

        \includegraphics[width=3in]{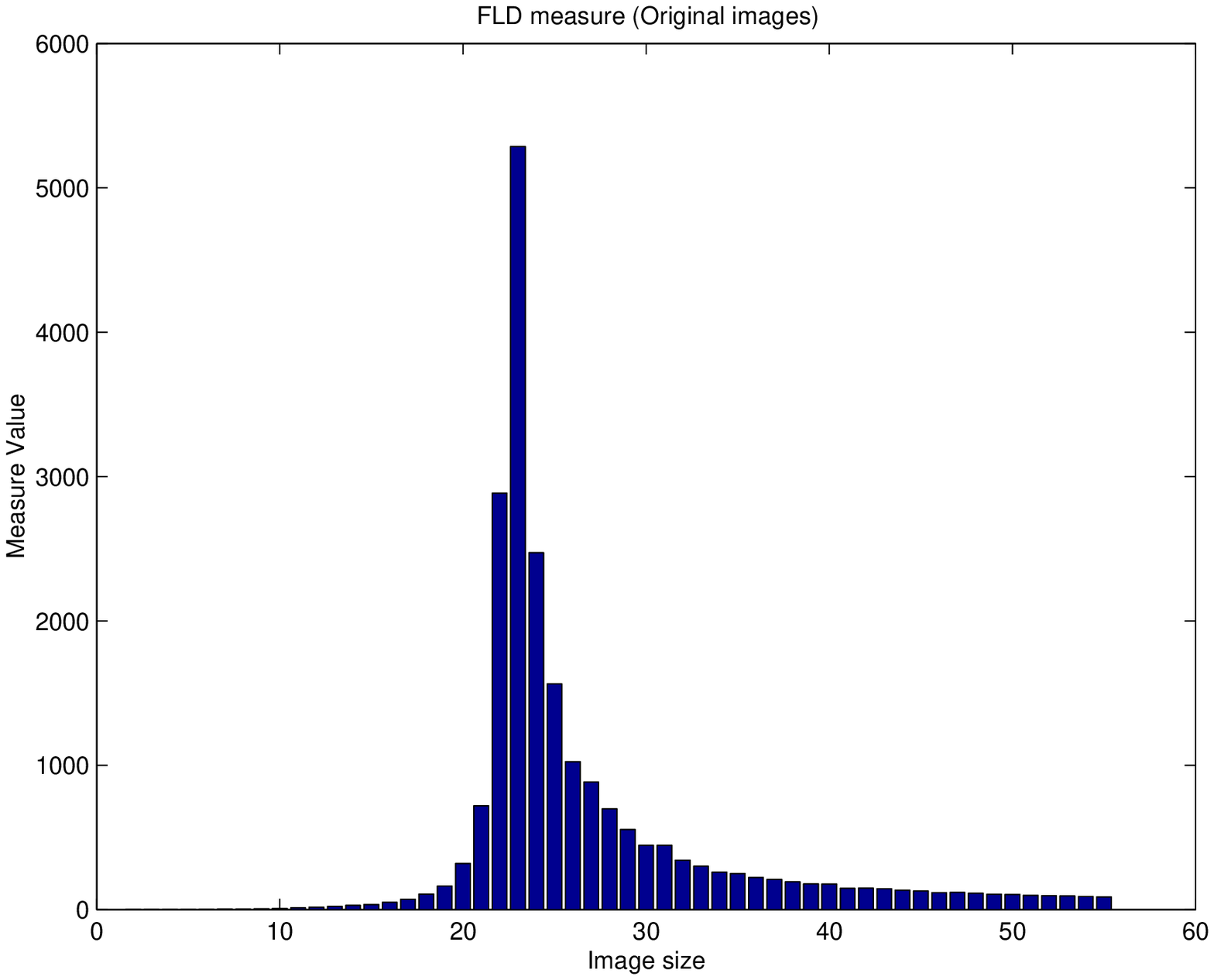}&
        \includegraphics[width=3in]{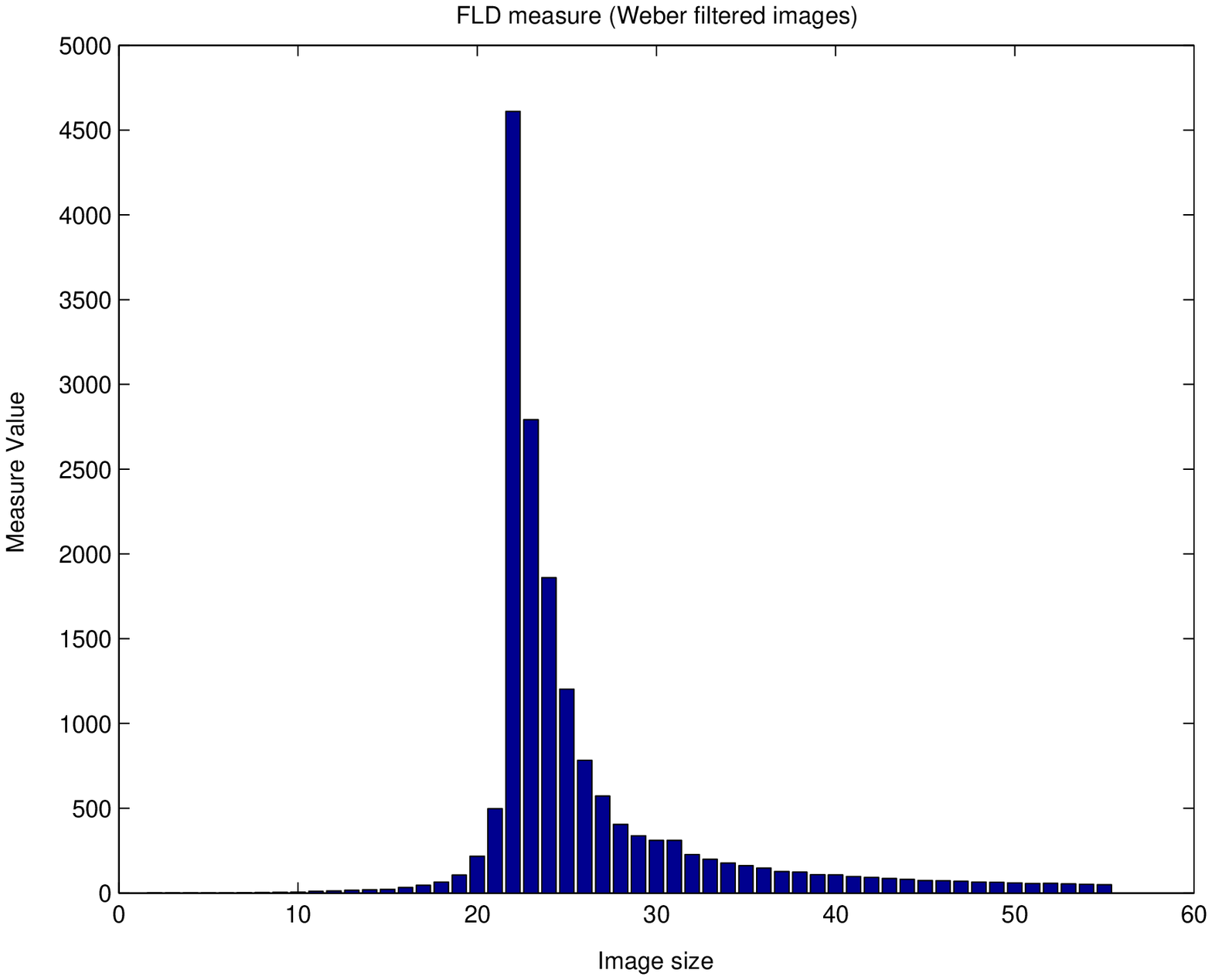}
        \end{tabular}
    \caption{FLD (Fisher Linear Discriminant Analysis) Measure.  Left plot shows results with
    the original images, right plot with Weber-filtered images.}
    \label{fig FLD}
\end{figure}

\begin{figure}
\centering
\begin{tabular}{cc}

        \includegraphics[width=3in]{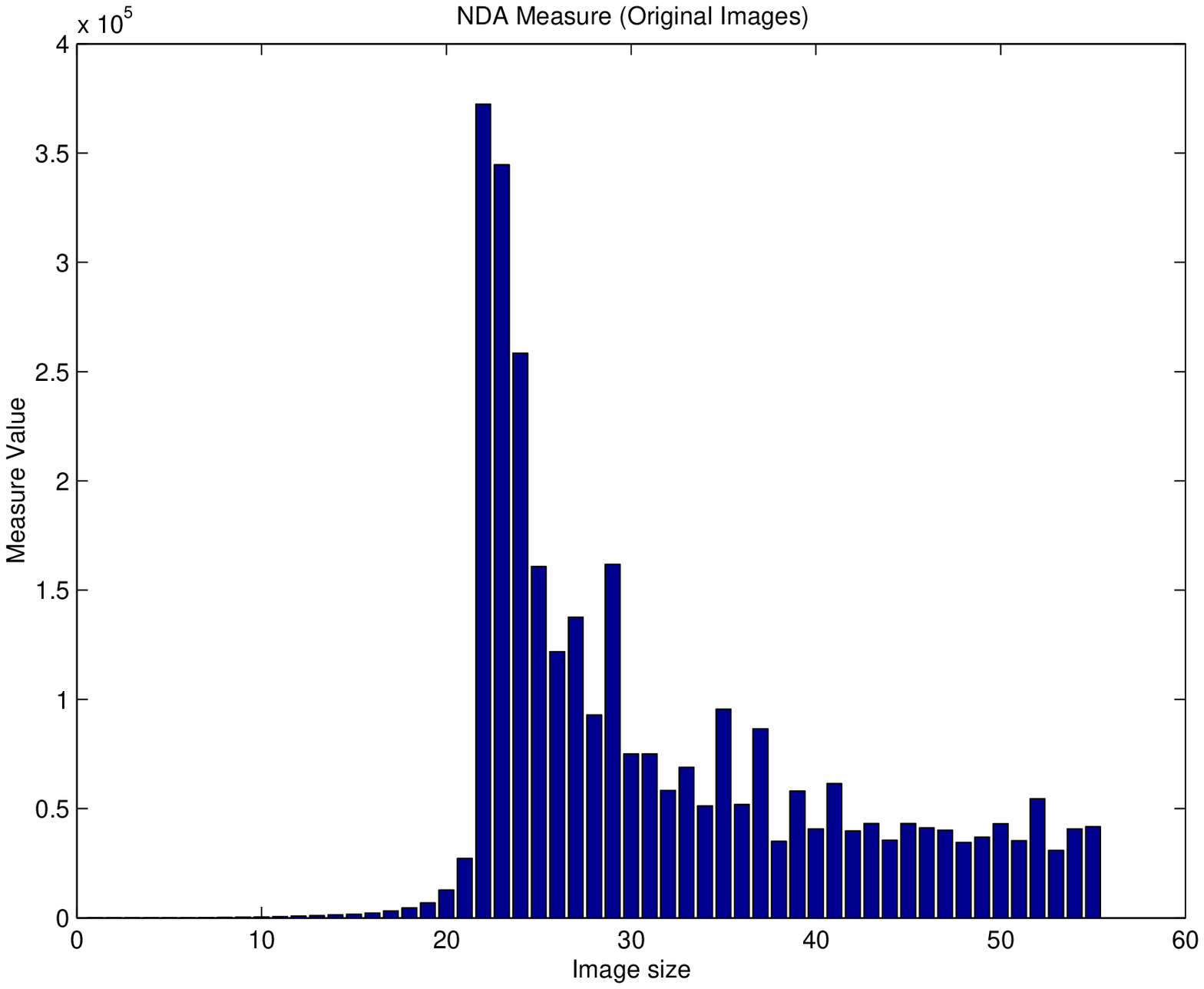}&
        \includegraphics[width=3in]{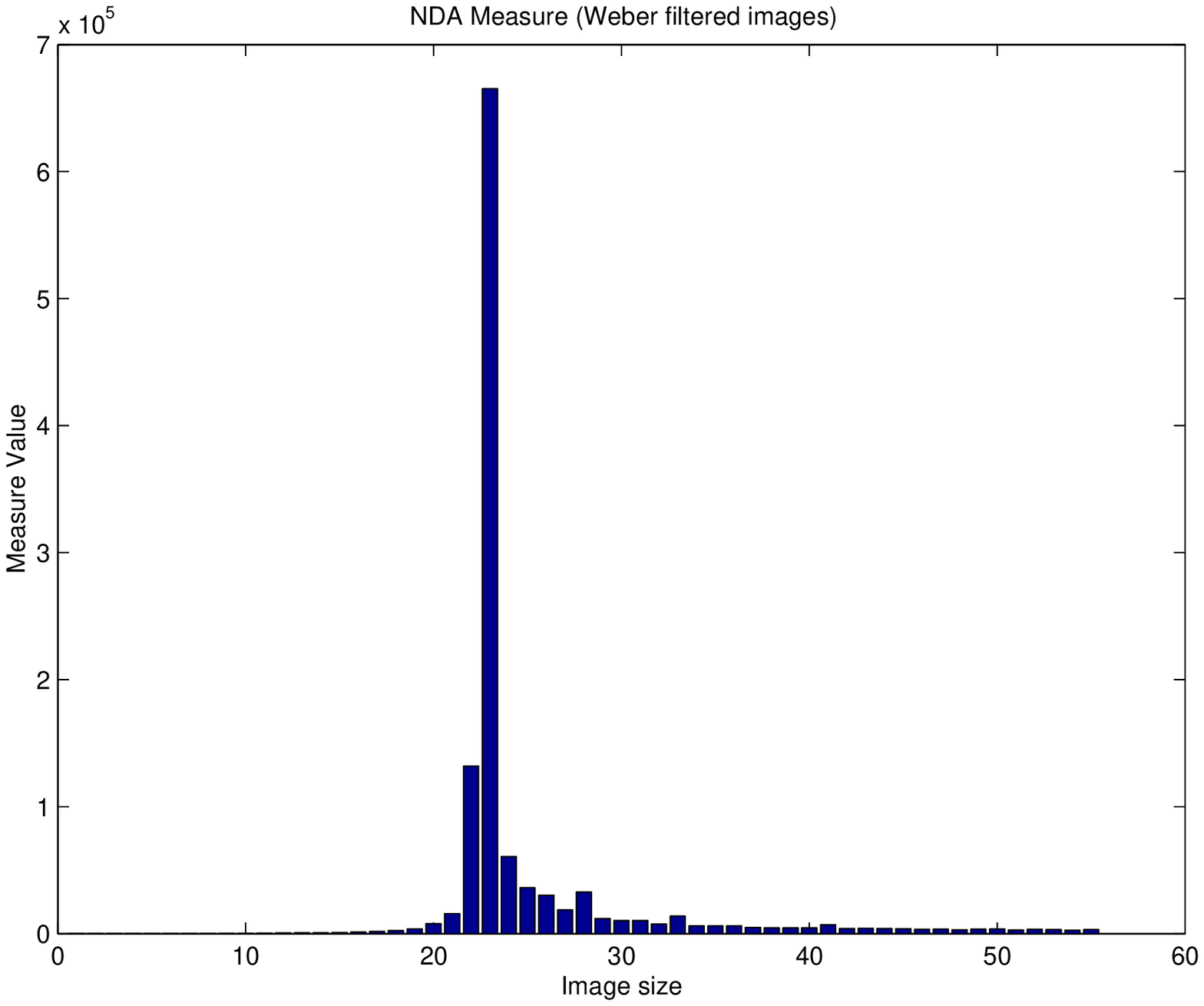}
        \end{tabular}
    \caption{NDA (non Parametric Discriminant Analysis) Measure.  Left plot shows results with
    the original images, right plot with Weber-filtered images.}
    \label{fig NDA}
\end{figure}

\begin{figure}
\centering
\begin{tabular}{cc}

        \includegraphics[width=2.8in]{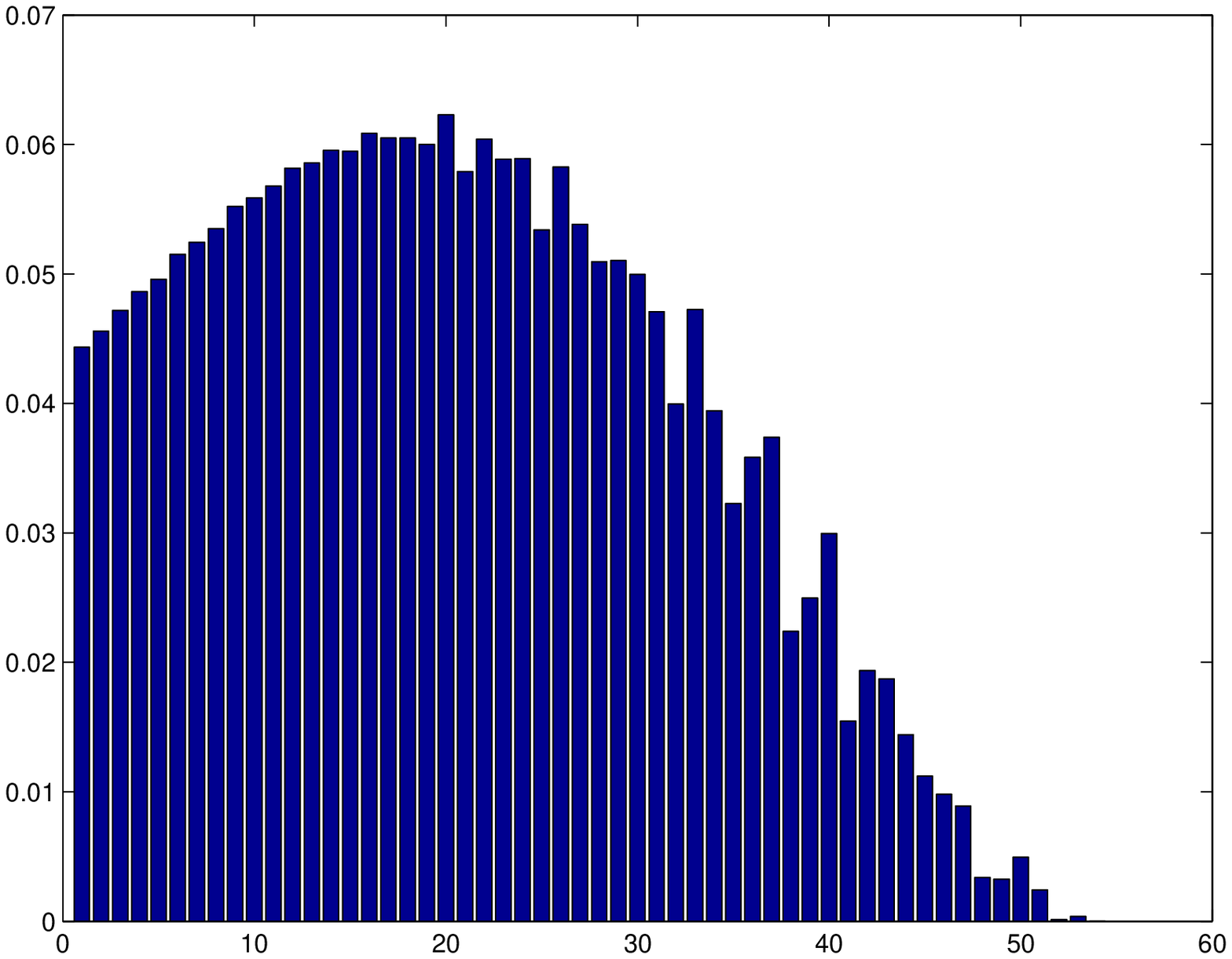}&
        \includegraphics[width=2.8in]{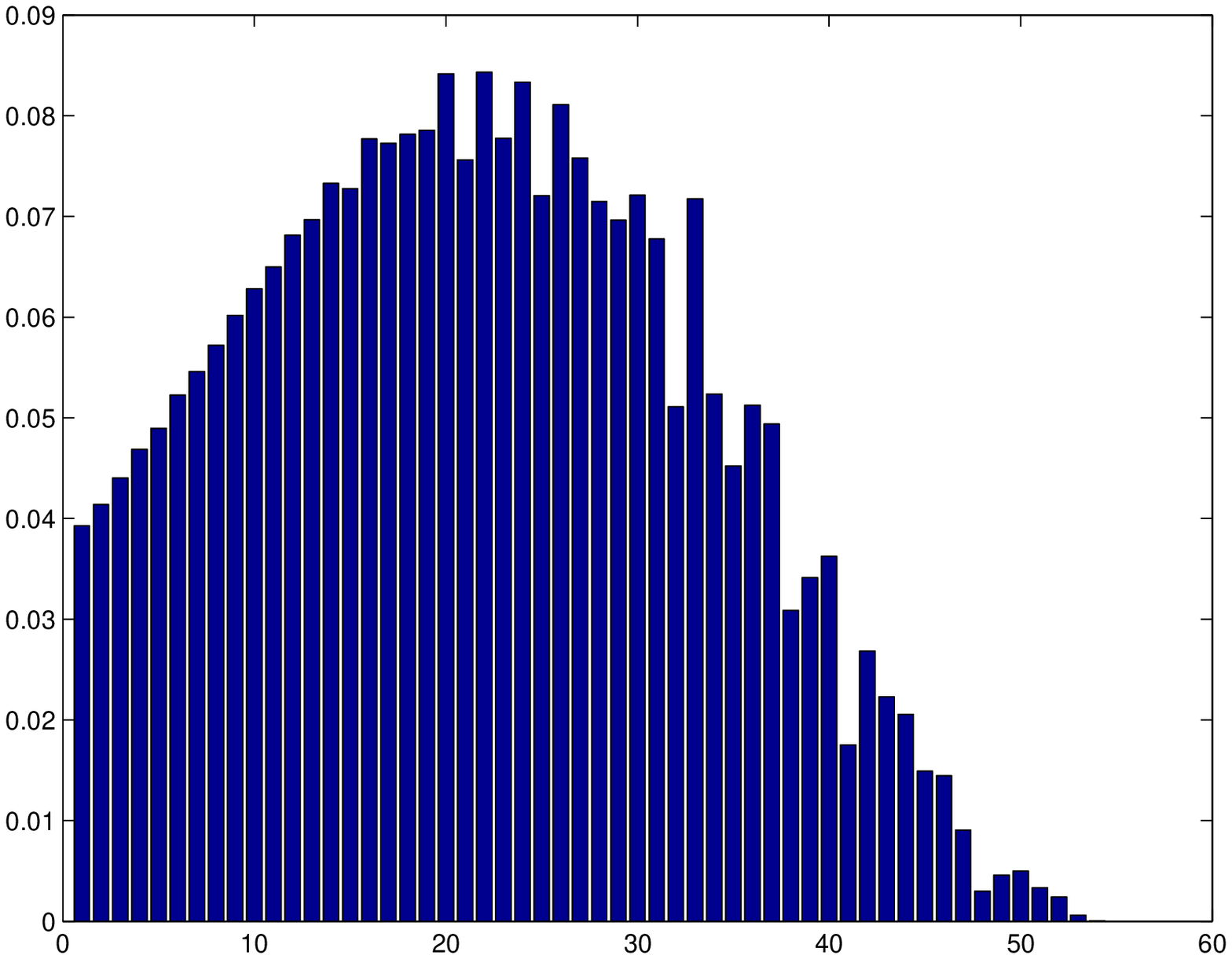}
        \end{tabular}
    \caption{MI (mutual information) Measure.  Left plot shows results with
    the original images, right plot with Weber-filtered images.}
    \label{fig MI}
\end{figure}

We used the three separability measures described in the previous section
(FLD, NDA and MI) for evaluating the recognition performance as a function
of image size.  To this end, $20$ subjects were randomly selected, with each
of the subjects having more than $25$ images to compute the different geometrical
and statistical measures described above.  All numerical experiments were carried
out with the original image set, and a second image set obtained by applying
Weber-filtering \cite{GrossBrajovic03}.  In this way we were able to address
how our results depend on the presence of low spatial frequencies in the
down-sized images.\\
Figures \ref{fig FLD}, \ref{fig NDA}, and \ref{fig MI} show the dependence of the FLD (Fisher
Linear Discriminant Analysis), NDA (non Parametric Discriminant Analysis), and MI (mutual
information) measures, respectively, on image size.  Each of the three measures reveals a distinct
maximum at approximately the same image size (around $22\times22$ pixels). As one can appreciate
from the sample images shown in Figure \ref{FRGC_database_small}, this image size translates to
roughly $8$ to $10$ cycles per face, what compares favorably to the psychophysical results as
described in the introduction. The recognition performance (in terms of discriminability) also
reveals some dependency on whether the original images are used, or whether the face images were
Weber-filtered. Specifically, the maxima show a trend to get more pronounced with Weber-filtering.
At the same time, the amplitudes of NDA and MI (but not FLD) grow, indicating an increased
recognition performance when only a small band of spatial frequencies is used.  This behavior of
our artificial face recognition system is also consistent with corresponding psychophysical
observations with humans -- the bandwidth for face recognition was estimated to be around two
octaves (e.g., \cite{Nasanen1999}).

\section{Summary and Conclusions}
Psychophysical studies suggest that for face recognition, human observers make use of a narrow band
at low spatial frequencies ($10$ to $15$ cycles per face, bandwidth two octaves).  Here, we
evaluated the recognition performance of an artificial face recognition system as a function of
image size. Recognition performance was measured by three different measures (Fisher Linear
Discriminant Analysis, non Parametric Discriminant Analysis, and mutual information), which all
indicated a performance maximum of the artificial system at an image size of about $22\times22$
pixels.  This corresponds to spatial frequencies at around $8-10$ cycles per face, thus comparing
well to the range of measured psychophysical data (although the psychophysical data are somewhat
underestimated).  We also found an effect of the presence of low spatial frequencies in the image.
Recognition performance seems to even increase when low spatial frequencies are suppressed by
Weber-filtering. In other words, decreasing the bandwidth of the spectrum of spatial frequencies in
the face images increases the recognition performance, at least when measured by non Parametric
Discriminant Analysis and mutual information.  Such behavior is again in line with the narrow band
of critical
spatial frequencies found psychophysically.\\
The present study furthermore lends further support to the findings of Keil \cite{KeilFace2006} in
that the stimuli (i.e., face images) provide the explanation of the preference of a narrow spatial
frequency band for both human and artificial face recognition.  As a consequence, artificial face
recognition systems should focus on these frequencies to achieve an optimal recognition performance
(in terms od class separability). Because this critical spatial frequencies correspond to small
image patches, a further advantage is an economic use of resources for both processing and storing
faces.
\section{Acknowledgment}
MSK is supported by a \emph{Juan de la Cierva} program from the Spanish government.  Further
support for this work was provided by a grant TIC2003-00654 from the Ministerio de Ciencia y
Tecnologia, Spain.









\end{document}